\documentclass[12pt,a4paper]{article}

\usepackage{graphicx}

\begin{document}

\title{On classical super-radiance in Kerr-Newman-anti-de Sitter
black holes}
\author{Elizabeth Winstanley\\
{\it {Department of Applied Mathematics,}}\\ 
{\it {The University of Sheffield,}}\\
{\it {Hicks Building, Hounsfield Road,}}\\ 
{\it {Sheffield. S3 7RH U.K. }}
\\gr-qc/0106032}

\date{\today}

\maketitle

\begin{abstract}
We study in detail the modes of a classical scalar field on a 
Kerr-Newman-anti-de Sitter (KN-AdS) black hole.      
We construct sets of basis modes appropriate to the two possible
boundary conditions (``reflective'' and ``transparent'') 
at time-like infinity, and consider whether super-radiance
is possible.
If we employ ``reflective'' boundary conditions, 
all modes are non-super-radiant.
On the other hand, for ``transparent'' boundary conditions, 
the presence 
of super-radiance depends on our definition of positive
frequency. 
For those KN-AdS black holes having a globally time-like Killing
vector, the natural choice of positive frequency leads to no
super-radiance.
For other KN-AdS black holes, there is a choice of positive
frequency 
which gives no super-radiance, but for other choices
there will, in general,
be super-radiance.
\end{abstract}

\section{Introduction}
\label{sec:intro}

Although the subject of quantum field theory on black hole
geometries in asymptotically flat space
has been widely researched over the past
thirty years, there remain some aspects which are not fully 
understood.  
Many detailed computations of the expectation values
of quantities such as the renormalized 
expectation value of the stress tensor
have been performed on static,
spherically symmetric geometries such as Schwarzschild,
but the greater complexity of such calculations
on rotating black hole space-times has prevented 
correspondingly exhaustive study. 
One reason for the greater computational difficulty in
rotating black holes is that the metric and other
geometric quantities depend on two, rather than one,
variable.
Another key factor for quantum field theory is the
presence of super-radiant modes, as a result
of the lack of a globally time-like Killing vector.
These need to be dealt with in a subtle manner
(see \cite{ottewill,frolov} for discussion of this point).

In recent years there has been a great increase
in interest in black holes in asymptotically 
anti-de Sitter space, due to the conjectured
correspondence between gravity in the bulk of
anti-de Sitter space and conformal field theory on its
boundary (the AdS/CFT correspondence - see, for example,
\cite{aharony} for a review and comprehensive list of
references).
In the light of this, it is natural to instigate
a study of quantum field theory on black holes geometries
which are asymptotically anti-de Sitter, instead of
asymptotically flat.
Such an investigation may deepen understanding, not
only of the AdS/CFT correspondence itself,
but also of the outstanding problems of quantum 
field theory on asymptotically flat black holes.
Quantum field theory on asymptotically anti-de Sitter black holes
is also attractive to study
since there are black holes in anti-de Sitter space 
which have a stable Hartle-Hawking state \cite{hp}, 
whereas in asymptotically
flat space black holes can only be in unstable equilibrium 
with a heat bath at the Hawking temperature. 

There has already been a great deal of work on 
(both rotating and static) black holes in anti-de Sitter space,
primarily the BTZ \cite{banados1,banados2} black holes
which live in a three-dimensional universe with a negative
cosmological constant.  
As well as their classical properties, the behaviour of quantum
fields on the BTZ black hole has been extensively studied 
(see, for example, \cite{dasgupta,keski,myung,mann}).
One advantage of working in three dimensions is that
many of the computational aspects are greatly simplified
(for example, it is possible to write the Green's function 
for a quantum field in the Hartle-Hawking state in closed form
\cite{lifschytz,shiraishi,steif}).
However, there has been more recent interest in higher-dimensional
black holes in anti-de Sitter space, particularly 
Kerr-Newman-anti-de Sitter (KN-AdS)
black holes in various dimensions \cite{caldarelli,hhtr,hr}.  
Work to date has concentrated on calculating those quantities
of particular interest for the AdS/CFT correspondence, 
primarily within a classical framework
(including the work in Refs. \cite{mann1,das,awad,awad1,dehghani}), 
and there has been little study of conventional
quantum field theory on these backgrounds
(see, for example, \cite{hemming} for work to date). 

Our purpose in the present article is to lay some 
foundations for a detailed study of quantum field theory
on four-dimensional KN-AdS black holes,
by considering the modes of a classical scalar
field on this background.
In particular, we wish to 
discover whether the super-radiant modes,
which cause so much difficulty in Kerr-Newman
space-time, still exist in KN-AdS
black holes.
If there is an absence of super-radiance for (at least some)
KN-AdS black holes, then it may be that quantum field
theory in these space-times is, in some ways,
simpler than for asymptotically
flat Kerr-Newman black holes, and one may hope to 
tackle such questions as the existence of a Hartle-Hawking
state (which does not exist for Kerr-Newman 
geometries \cite{frolov,ottewill1,kay}).
The present article, however, concentrates on
purely classical scalar field modes, a detailed
understanding of which is an essential precursor
to serious study of quantum field theory effects
(for example, for calculating the semi-classical
entropy of the black hole \cite{ho,lee}).
We consider a scalar field as the simplest bosonic
field, since for Kerr-Newman black holes in asymptotically
flat space, super-radiance occurs for bosonic 
but not fermionic fields.
However, the quantum analogue of super-radiance does occur
for both bosonic and fermionic fields \cite{unruh}. 
For reviews and references to the original literature 
on classical super-radiance in Kerr-Newman black holes,
the reader is referred to \cite{dewitt} for a scalar fields, 
and \cite{chandrasekhar} for spin 1/2 fields, electromagnetic
and gravitational perturbations.

We begin, in section \ref{sec:geom}, by reviewing
those aspects of the geometry of KN-AdS which 
are useful for our purposes.
One interesting result at this stage is that 
there are KN-AdS black holes which possess 
a Killing vector which is time-like everywhere
outside the event horizon.
This suggests that there may be KN-AdS geometries
on which super-radiance is absent.
In section \ref{sec:modes} we separate the
wave equation for scalar fields and study, in particular,
the radial equation, by means of a new radial co-ordinate
$R$ which turns out to be particularly useful.
The next two sections 
tackle the more subtle issues of boundary conditions
and positive frequency, and 
we construct linearly independent sets of basis modes 
which will be useful when we return, in a subsequent
publication, to the subtle issue of constructing
suitable quantum vacuum states.
In section \ref{sec:super}
we reach our conclusions about the presence 
(or absence) of super-radiant modes.
Due to the time-like nature of infinity ${\cal {I}}$ in KN-AdS,
there is a choice of two types of boundary condition at 
infinity  \cite{avis}: ``reflective'' and ``transparent''.
For ``reflective'' boundary conditions, no particle flux escapes
through infinity and super-radiance is absent.
This is in agreement with an energy argument given by 
Hawking and Reall \cite{hr}.
For ``transparent'' boundary conditions, whether or not there
are super-radiant modes depends on our prescription of
positive frequency.
For those KN-AdS black holes having a globally time-like 
Killing vector, the natural definition of positive frequency
leads only to non-super-radiant modes.
Otherwise, the choice of positive frequency is ambiguous,
with the analogue of the conventional choice in Kerr-Newman
black holes leading to super-radiant modes (although there
is another choice of positive frequency in which all modes
are non-super-radiant, whereas in Kerr-Newman super-radiance
is inevitable, whatever the choice of positive frequency).
Section \ref{sec:conc} contains our thoughts on the consequences
of these results for quantum field theory.

\section{Geometric structure of KN-AdS black holes}
\label{sec:geom}

In Boyer-Lindquist-like co-ordinates the metric of the
KN-AdS  black hole takes the form \cite{carter}
\begin{eqnarray}
  ds^{2} & = & 
  -\frac {\Delta _{r}}{\rho ^{2}} \left[ 
  dt - \frac {a \sin ^{2} \theta }{\Sigma } d\phi \right] ^{2}
  + \frac {\rho ^{2}}{\Delta _{r}} dr ^{2} 
  + \frac {\rho ^{2}}{\Delta _{\theta }} d\theta ^{2}  
\nonumber \\ & &   
  + \frac {\Delta _{\theta }\sin ^{2} \theta }{\rho ^{2}} \left[
  a \, dt - \frac {r^{2}+a^{2}}{\Sigma } d\phi \right] ^{2},
\label{eq:metric}
\end{eqnarray}
where 
\begin{eqnarray*}
  \rho ^{2} & = & r^{2}+a^{2} \cos ^{2} \theta , 
\\
  \Sigma & = & 
\begin{displaystyle} 
1- \frac {a^{2}}{l^{2}} , 
\end{displaystyle}
\\
  \Delta _{r} & = &
\begin{displaystyle}
 \left( r^{2}+ a^{2} \right) \left(
    1 + \frac {r^{2}}{l^{2}} \right)
    - 2Mr + z^{2} , 
\end{displaystyle}
\\
  \Delta _{\theta } & = & 
\begin{displaystyle}
1 -\frac {a^{2}}{l^{2}} \cos ^{2} \theta .
\end{displaystyle}
\end{eqnarray*}
Throughout this paper the metric has signature $(-,+,+,+)$ and
we use units in which $G=c=1$.
The parameter $a$ describes the rotation (and vanishes for
the Schwarzschild-AdS black hole), $M$ is the mass parameter
and $z^{2}$ represents
the total squared charge (i.e. the sum of the electric
charge squared and the magnetic charge squared when both 
are present).
This metric is a solution of the Einstein-Maxwell equations
with a negative cosmological constant $\Lambda = -3/l^{2}$,
and is valid only for $a^{2}<l^{2}$.
The singular limit in which $a^{2}\rightarrow l^{2}$
has been considered by, for example, \cite{hhtr}.

For values of the mass parameter $M$ above a critical value
$M_{c}$, 
the function $\Delta _{r}$ has two roots: $r=r_{+}$, which 
is the location of the outer (event) horizon, and $r=r_{-}$.
Given that $\Delta _{r}$ is a quartic, there is no simple
closed form expression for $r_{\pm }$, although it is
straightforward to show that $\Delta _{r}$ has only two
positive real roots. 
If $M<M_{c}$, then $\Delta _{r}$ has no zeros and
there is a naked singularity, whilst for
$M=M_{c}$, we have an extremal black hole. 
In this paper we shall assume that $M>M_{c}$ and
concentrate on the geometry of the black hole
exterior to the event horizon $r=r_{+}$.
The event horizon has angular velocity $\omega _{+}$ given by
\begin{displaymath}
  \omega _{+} = \frac {a\Sigma }{r_{+}^{2}+a^{2}} .
\end{displaymath}

These Boyer-Lindquist-like
co-ordinates are the most convenient in which to find 
the scalar field mode solutions in the next section.
It is clear from the
form of the metric (\ref{eq:metric}), how the Kerr-Newman
geometry arises as the limit $l \rightarrow \infty $.
However, the use of Boyer-Lindquist-like co-ordinates
obscures the structure of the geometry at infinity.
At infinity, the metric (\ref{eq:metric}) describes
anti-de Sitter space as seen by a rotating observer.
This can most readily be seen by making the co-ordinate
transformation \cite{henneaux}:
\begin{displaymath}
  T=t, \qquad
  \Phi = \phi + \frac {a}{l^{2}} t, \qquad
  y \cos \Theta = r \cos \theta, \qquad
  y^{2} = \frac {1}{\Sigma } \left[ 
    r^{2} \Delta _{\theta } + a^{2} \sin ^{2} \theta \right] .
\end{displaymath}
A long but straightforward calculation shows that the most
compact form for the metric components in these co-ordinates is:
\begin{eqnarray}
  g_{TT} & = & 
\begin{displaystyle}
-1 - \frac {y^{2}}{l^{2}} -
    \frac {\delta \Delta _{\theta }^{2}}{\rho ^{2} \Sigma ^{2} },
\end{displaystyle}
\nonumber \\
  g_{T\Phi } & = & 
\begin{displaystyle}
\frac {a \delta \Delta _{\theta }}{\rho ^{2} 
    \Sigma ^{2} } \sin ^{2} \theta ,
\end{displaystyle}
\nonumber \\  
  g_{\Phi \Phi } & = & 
\begin{displaystyle}
y^{2} \sin ^{2} \Theta -
    \frac {a^{2} \delta }{\rho ^{2} \Sigma ^{2}} \sin ^{4} \theta ,
\end{displaystyle}
\nonumber \\  
  g^{yy} & = & 
\begin{displaystyle}
1 + \frac {y^{2}}{l^{2}} +
    \frac {\delta \Delta _{\theta }^{2} r^{2}}{\rho ^{2}
      \Sigma ^{2}y^{2}} ,
\end{displaystyle}
\nonumber \\
  g^{y\Theta } & = & 
\begin{displaystyle}
- \frac {a^{2} \delta \Delta _{\theta } r}{
    \rho ^{2} \Sigma ^{3/2} y^{3} (r^{2}+a^{2})^{1/2}} 
      \sin \theta \cos \theta ,
\end{displaystyle}
\nonumber \\  
  g^{\Theta \Theta } & = & 
\begin{displaystyle}
\frac {1}{y^{2}} +
    \frac {a^{4} \delta }{\rho ^{2} \Sigma y^{4} (r^{2}+a^{2})}
      \sin ^{2} \theta \cos ^{2} \theta , 
\end{displaystyle}
\label{eq:varmetric}
\end{eqnarray}
where
\begin{displaymath}
\delta = -2Mr + z^{2} ,
\end{displaymath}
and $r$ and $\theta $ are functions of $y$ and $\Theta $.
In these co-ordinates it becomes apparent that in the
case $M=0$, $z=0$ (when $\delta =0$), 
the metric (\ref{eq:metric}) is simply
anti-de Sitter space as seen by a rotating observer, and also that,
as $r\rightarrow \infty $ (which corresponds to 
$y \rightarrow \infty $) the geometry approaches
anti-de Sitter space.  
The $(T,y,\Theta ,\Phi )$ co-ordinates are therefore more
physical and we will refer to them especially for
interpreting the modes at infinity.
However, due to the greater complexity of the metric
(\ref{eq:varmetric}), it will be easier to perform 
computations using $(t,r,\theta ,\phi )$ co-ordinates
and then transform to $(T,y,\Theta, \Phi )$ co-ordinates.
The angular velocity of the event horizon in $(T,y,\Theta ,\Phi )$
co-ordinates is
\begin{displaymath}
\Omega _{+} = \omega _{+} + \frac {a}{l^{2}}
= \frac {a(1+r_{+}^{2}l^{-2})}{r_{+}^{2}+a^{2}}.
\end{displaymath}

In a rotating black hole space-time, there are two 
surfaces (apart from the event horizon) of importance,
especially for quantum field theory.
They are the stationary limit surface
and the velocity of light surface. 
The stationary limit surface is the surface
where an observer can no longer remain
stationary with respect to infinity,
but is forced to rotate by the dragging
of inertial frames near the event horizon.
For the KN-AdS black hole, the stationary
limit surface will be the surface on which the 
Killing vector $\eta =\partial /\partial T$ (which is 
time-like at infinity) becomes null.
Using $(T,y,\Theta ,\Phi )$ co-ordinates,
it is straightforward to show that $g_{TT}>0$ at the 
event horizon, so that $\eta $ is space-like close to the
event horizon.
Therefore,  for all KN-AdS black holes, there
has to be a stationary limit surface and ergosphere.

The Killing vector generating the event horizon
is
\begin{displaymath}
  \zeta = \frac {\partial }{\partial T} + \Omega _{+}
    \frac {\partial }{\partial \Phi },
\end{displaymath}
and the velocity of light surface (if it exists) is the
surface outside the event horizon on which this
vector becomes null.
Outside the velocity of light surface, it is impossible
for an observer to rigidly rotate with the same
angular velocity as the event horizon.
As $y\rightarrow \infty $,
\begin{displaymath}
  g_{\tau \nu } \zeta ^{\tau } \zeta ^{\nu } 
    = -1 -\frac {y^{2}}{l^{2}} + \Omega _{+}^{2} y^{2}
    \sin ^{2} \Theta +O(y^{-1}).
\end{displaymath}
Therefore, if $\Omega _{+}^{2}<l^{-2}$, the Killing vector
$\zeta $ is time-like at infinity and there is no velocity of
light surface \cite{hr}.  
For these black holes possessing a globally time-like Killing
vector, as we shall discuss in section \ref{sec:super}, 
there is a natural definition of positive frequency 
for our classical scalar field modes,
and we may anticipate the absence of super-radiance.

Our purpose in the present article is to compare
the behaviour of classical scalar field modes in 
KN-AdS as opposed to ordinary Kerr-Newman black holes.
It is therefore useful to sketch the Penrose diagram
for the geometry outside the event horizon
on the axis of symmetry, which 
is shown in figure 1 for both the KN-AdS  and 
asymptotically flat Kerr-Newman black holes (cf. Ref. \cite{carter1}).
We also sketch the Penrose diagram for 
(the covering space of) anti-de Sitter
space itself for comparison.
\begin{figure}
\begin{center}
\includegraphics[height=3in]{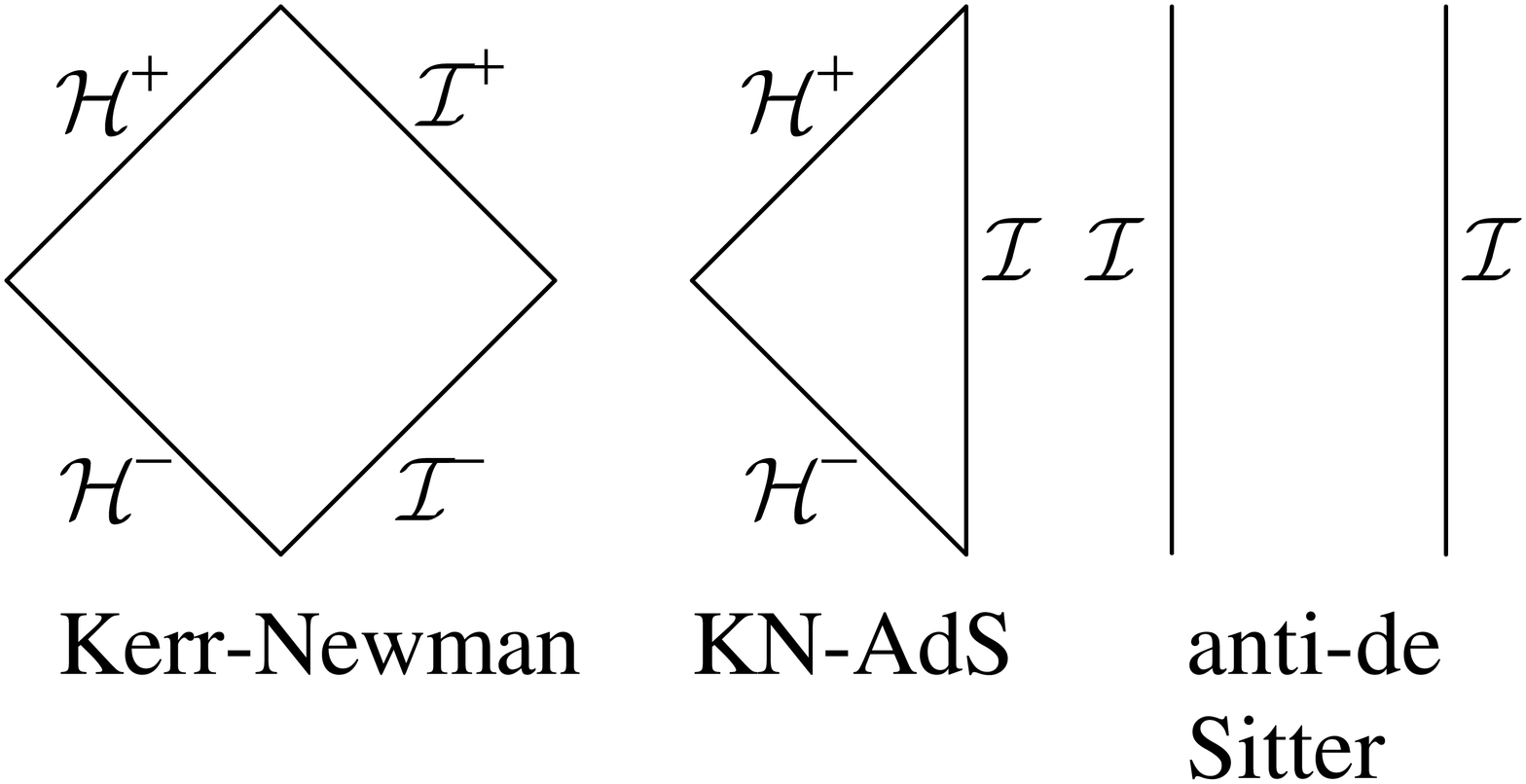}
\caption{Penrose diagrams for the geometry outside the
event horizon of Kerr-Newman and KN-AdS black holes,
along the axis of symmetry. We also
show the Penrose diagram for
(the covering space of) anti-de Sitter space for
comparison.}
\label{figure1}
\end{center}
\end{figure}
In the Penrose diagrams, ${\cal {H}}^{\pm }$ denote the future
and past event horizons, whilst ${\cal {I}}^{\pm }$ are future
and past null infinity for the Kerr-Newman black hole.
For the KN-AdS black hole and anti-de Sitter space, 
infinity is time-like and denoted simply ${\cal {I}}$.
Corresponding Penrose diagrams for BTZ black holes in
three-dimensional anti-de Sitter space may be found in
\cite{banados2,lifschytz,kim}, and 
for topological black holes in anti-de Sitter space in \cite{mann2}.

\section{Classical scalar field modes}
\label{sec:modes}

In this section we construct classical field modes
for a scalar field with general coupling to the curvature.
As is well known, for Kerr-Newman black holes in asymptotically
flat space, classical super-radiance effects occur only 
for bosonic fields \cite{dewitt,chandrasekhar}, 
although quantum super-radiance occurs
for both bosonic and fermionic fields \cite{unruh}.
Since our interest in this paper is classical super-radiance,
we therefore need only consider a bosonic field, and 
we focus on a scalar field for simplicity.
It is expected that the scalar field will exhibit all
the subtleties associated with both the black hole's
rotation and the asymptotic structure of the space-time.

Therefore, we consider a general scalar field $\Psi $ 
satisfying the equation
\begin{equation}
  \left[ \nabla _{\nu } \nabla ^{\nu } - \xi R
 - \mu ^{2} \right] \Psi = 0,
\label{eq:KGone}
\end{equation}
where $R$ is the Ricci scalar of the geometry, $\xi $ is a 
coupling constant and $\mu $ is the mass of the field.
At this stage we make no assumptions about the mass $\mu $ 
or the value of the coupling constant $\xi $.
For the KN-AdS geometry, $R=-12/l^{2}$, so the equation
(\ref{eq:KGone}) takes the form
\begin{equation}
  \left[ \nabla _{\nu } \nabla ^{\nu } - {\tilde {\mu }}^{2}
    \right] \Psi =0
\label{eq:KGtwo}
\end{equation}
where
\begin{displaymath}
  {\tilde {\mu }}^{2}=\mu ^{2}-\frac {12\xi }{l^{2}}
\end{displaymath}
is an effective ``mass squared'' for the scalar field
(although ${\tilde {\mu }}^{2}$ need not be positive, for
example, for conformally coupled massless fields, $\mu =0$
and $\xi =1/6$, so that ${\tilde {\mu }}^{2}=-2/l^{2}<0$).
Note that, due to the constancy of the curvature
of the geometry, models with different values of
$\xi $ and $\mu $ can give rise to the same field equation
(\ref{eq:KGtwo}).
This ambiguity was considered in 
\cite{breitenlohner1,breitenlohner2} for 
scalar fields in super-gravity on anti-de Sitter space.

The wave equation (\ref{eq:KGtwo}) separates in the Boyer-Lindquist-like
co-ordinates $(t,r,\theta ,\phi )$, so we consider solutions
of the form
\begin{equation}
  \Psi (t,r,\theta , \phi )= e^{-i\omega t} e^{im\phi }
    F_{\omega lm}(r) G_{\omega lm}(\theta ).
\label{eq:modes}
\end{equation}
For field modes in anti-de Sitter space, regularity of the
modes at the origin implies that the frequency $\omega $
will take discrete values \cite{avis}.
For KN-AdS black holes, however, we do not need to impose
regularity at the origin and so $\omega $ is continuous.
The angular quantum number $m$ is required to be an integer,
and $l$ is an integer labelling the angular eigenfunctions.
When studying quantum field theory on Kerr-Newman black holes,
it is conventional at this stage to define ``positive frequency''
by taking $\omega >0$ \cite{ottewill}.
However, at the moment we shall not fix the sign of $\omega $,
and shall postpone a discussion of the appropriate definition of
positive frequency until section \ref{sec:super}.

The separated modes (\ref{eq:modes})
give the following radial and angular equations:
\begin{eqnarray}
  0 & = &
    \Delta _{r} \partial _{r}
    \left( \Delta _{r} \partial _{r} F_{\omega lm}(r) \right) 
\nonumber
\\ & & +
    \left\{ \left[ \left( r^{2}+a^{2} \right) \omega - ma \Sigma 
    \right] ^{2} - K_{\omega lm} \Delta _{r} - r^{2} 
    {\tilde {\mu }}^{2} \Delta _{r} \right\} F_{\omega lm}(r) ,
\nonumber
\\
  0 & = & 
    \Delta _{\theta } \sin \theta \partial _{\theta }
    \left( \Delta _{\theta }
    \sin \theta \partial _{\theta } 
    G_{\omega lm}(\theta ) \right) 
\label{eq:angular}
\\ & & +
    \left\{ K_{\omega lm} \Delta _{\theta } \sin ^{2} \theta 
    - \left( a\omega \sin ^{2} \theta - m \Sigma \right) ^{2} 
    - a^{2} {\tilde {\mu }}^{2} 
    \Delta _{\theta } \sin ^{2} \theta \cos ^{2} \theta
    \right\} G_{\omega lm},
\nonumber
\end{eqnarray}
where $K_{\omega lm}$ is a separation constant.
Due to the $\Delta _{\theta }$ terms, the angular equation
(\ref{eq:angular}) is more complex that that for the usual
spheroidal harmonic functions \cite{abramowitz}.
However, imposing regularity of $G_{\omega lm }(\theta )$ 
at $\theta =0$ and $\pi $ will give a discrete sequence of
real eigenvalues $K_{\omega lm}$ which we label by the integer
$l=0,1,\ldots $ (which, unlike the case in asymptotically
flat space, is not necessarily related to the angular
quantum number $m$).
We shall take the corresponding eigenfunctions to be real
(as is the case for the spheroidal harmonics).
Given a lack of a simple formula for the eigenvalues 
and eigenfunctions for ordinary spheroidal harmonics,
we do not anticipate such a formula in the present situation.
We will not consider the angular equation further in the 
present work, and shall focus on the radial equation.
However, detailed study of the angular equation will be
necessary for any complete computation of quantum field
theory expectation values in KN-AdS.

The usual ``tortoise'' co-ordinate $r^{*}$ for KN-AdS is
defined by the equation
\begin{displaymath}
  \frac {dr^{*}}{dr} = \frac {r^{2}+a^{2}}{\Delta _{r}}.
\end{displaymath}
If we further define a new radial function 
${\tilde {F}}_{\omega lm }$ by
\begin{displaymath}
  F_{\omega lm }(r)=\left( r^{2}+a^{2} \right) ^{-1/2} 
  {\tilde {F}}_{\omega lm }(r^{*}),
\end{displaymath}
then ${\tilde {F}}_{\omega lm }(r^{*})$ 
satisfies the differential equation
\begin{displaymath}
  \partial ^{2}_{r^{*}} {\tilde {F}}_{\omega lm} (r^{*}) 
  + {\tilde {V}}_{\omega lm }(r^{*})
  {\tilde {F}}_{\omega lm} (r^{*}) =0,
\end{displaymath}
where the potential ${\tilde {V}}_{\omega lm }(r^{*})$ is given by
\begin{eqnarray}
  {\tilde {V}}_{\omega lm} (r^{*}) & = &
    \left[ \omega - \frac {ma\Sigma }{r^{2}+a^{2}} \right] ^{2}
    -\frac {3a^{2} \Delta _{r}^{2}}{(r^{2}+a^{2})^{4}}     
    -\frac {\Delta _{r}}{(r^{2}+a^{2})^{2}} 
    \left( K_{\omega lm } + r^{2}{\tilde {\mu }}^{2} \right)
\nonumber
\\ & &     
    +\frac {\Delta _{r}}{(r^{2}+a^{2})^{3}} 
    \left( 2a^{2} - 2Mr + 2z^{2} -\frac {2r^{4}}{l^{2}} \right) ,
\label{eq:Vtilde}
\end{eqnarray}
and $r$ is to be considered as a function of $r^{*}$.
The variable $r^{*}$ is particularly useful for studying the 
structure of the geometry close to the event horizon (and constructing
the Penrose diagram given in figure 1).  
At the event horizon, 
\begin{equation}
  {\tilde {V}}_{\omega lm }\sim 
  {\tilde {\omega }}^{2} = \left( \omega - m \omega _{+} \right) ^{2},
\label{eq:tildeomega}
\end{equation}
so we get the usual ingoing and outgoing field modes at
the event horizon:
\begin{equation}
  {\tilde {F}}_{\omega lm } (r^{*}) \sim 
    e^{\pm i {\tilde {\omega }} r^{*}} .
\label{eq:r*hor}
\end{equation}

However, at infinity $r\rightarrow \infty $, the ``tortoise''
co-ordinate $r^{*}$ approaches a finite value $r^{*}_{\infty }$.
Further, the potential ${\tilde {V}}_{\omega lm }$ diverges
at infinity unless ${\tilde {\mu }}^{2}=-2/l^{2}$.  
Some studies of the quasi-normal modes on asymptotically anti-de Sitter
black holes (see, for example, 
\cite{horowitz,cardoso,cardoso1,wang1,wang2,wang3}) 
have considered
massless, minimally coupled scalar fields, for which
${\tilde {\mu }}=0$.
Therefore, in that case, the potential is divergent at infinity
and the boundary condition ${\tilde {F}}_{\omega lm }\rightarrow 0$
as $r^{*}\rightarrow r^{*}_{\infty }$ was employed.
For massless, conformally coupled scalar fields, 
whose quasi-normal modes were studied in \cite{chan1,chan2}
(see \cite{mann2} for a summary of this work), $\xi = 1/6$ and
$\mu =0$, so that ${\tilde {\mu }}^{2} =-2 / l^{2}$ and
the potential (\ref{eq:Vtilde}) remains finite at infinity.
As commented in \cite{chan1,chan2}, in this case the potential most closely
resembles that for black holes in asymptotically flat space.  

In order to study the wave-like properties of the field modes
at infinity, it is helpful to introduce another radial 
variable, $R$, which tends to infinity at both the event horizon
and infinity, giving an infinite variable range.
The simplest such co-ordinate is:
\begin{displaymath}
  R=\log \left( r-r_{+} \right) .
\end{displaymath}
If we define a new function $f_{\omega lm }(R)$ by
\begin{equation}
  F_{\omega lm }(r) = \left( r- r_{+} \right) ^{1/2}
    \Delta _{r} ^{-1/2} f_{\omega lm }(R),
\label{eq:fdef}
\end{equation}
then the governing differential equation is now
\begin{equation}
\partial _{R}^{2} f_{\omega lm }(R)
+ V_{\omega lm } (R) f_{\omega lm }(R) = 0,
\label{eq:Rradial}
\end{equation}
where
\begin{eqnarray*}
  V_{\omega lm }(R) & = & 
    \left( r-r_{+} \right) ^{2} \Delta _{r}^{-2}
    \left[ \left( r^{2}+a^{2} \right) \omega - ma \Sigma \right] ^{2}
    - K_{\omega lm } \left( r-r_{+} \right) ^{2} \Delta _{r}^{-1}
\\   & &  
    -{\tilde {\mu }}^{2} r^{2} \left( r-r_{+} \right) ^{2}
    \Delta _{r}^{-1}
    -\frac {1}{4} + \frac {1}{4} \left( r-r_{+} \right) ^{2}
    \Delta _{r}^{-2} \left( \frac {d\Delta _{r}}{dr} \right) ^{2}
\\ & &     
    -\frac {1}{2} \left( r-r_{+} \right) ^{2} \Delta _{r}^{-1}
    \frac {d^{2}\Delta _{r}}{dr^{2}} .
\end{eqnarray*}
Near the event horizon, as $R\rightarrow -\infty $,  
\begin{equation}
  V_{\omega lm } \sim \left( \frac {d\Delta _{r}}{dr} \right) ^{-2}
  \left( r_{+}^{2} + a^{2} \right) ^{2} 
  {\tilde {\omega }}^{2} =  \Omega ^{2},
\label{eq:Omegadef}
\end{equation}
giving $f_{\omega lm } \sim e^{\pm i \Omega R}$, in agreement
with the ingoing and outgoing waves (\ref{eq:r*hor}).
At infinity, $R\rightarrow \infty $, the potential $V_{\omega lm }$
remains finite for all values of the coupling constants, which
is the main advantage of this choice of co-ordinate.
We have
\begin{displaymath}
  V_{\omega lm } \sim - \frac {9}{4} -{\tilde {\mu }}^{2} l^{2} .
\end{displaymath}
The form of the mode solutions at infinity therefore depends
on the effective ``mass-squared'' ${\tilde {\mu }}^{2}$ (which,
recall, need not be positive).
If ${\tilde {\mu }}^{2}< -9/4l^{2}$ (hereafter referred to 
as ``case 1''), then 
$V_{\omega lm }\sim +q^{2}$ at infinity, where $q$ is real,
and 
\begin{displaymath}
  f_{\omega lm }(R) \sim e^{\pm iqR} .
\end{displaymath}
On the other hand, if ${\tilde {\mu }}^{2} > -9/4l^{2}$,
(``case 2'') we have $V_{\omega lm } \sim -q^{2}$ (with $q$ again real), 
so that
\begin{displaymath}
  f_{\omega lm }(R) \sim e^{\pm qR}.
\end{displaymath}

At this stage it is useful to compare the form of our
field modes with those studied by other authors on
anti-de Sitter space \cite{avis}.
There has recently been a revival in the study of
field modes on anti-de Sitter space due to the
AdS/CFT correspondence, according to which field
modes in the bulk correspond to either source
terms or states in the boundary theory, depending
on whether the modes are non-normalizable or
normalizable, respectively \cite{keski,balasubramanian,chang}.
In case 1, the radial function $F_{\omega lm }(r)$ behaves
like $r^{-3/2}$ at infinity, and all modes
are well-behaved at infinity and normalizable.
In case 2, the behaviour of the radial function 
at infinity is:
\begin{equation}
  F_{\omega lm }^{\pm } \sim r^{-3/2\pm q}
\label{eq:infinity}
\end{equation}
where
\begin{displaymath}
  q^{2}=\frac {9}{4}+{\tilde {\mu }}^{2} l^{2}>0 .
\end{displaymath}
If ${\tilde {\mu }}<0$, then both modes
tend to zero at infinity, although only $F^{-}_{\omega lm }$
is normalizable if $q>1/2$ (which corresponds to 
${\tilde {\mu }}^{2} l^{2}>-2$).
If ${\tilde {\mu }}\ge 0$, then only $F^{-}_{\omega lm }$ 
vanishes at infinity, the other mode diverging. 
It can be checked that the leading behaviour (\ref{eq:infinity})
is in agreement with that of the modes in \cite{avis}
for massless conformally coupled scalar fields (when 
${\tilde {\mu }}^{2} l^{2} = -2$).
For massless, minimally coupled, fields (such
as considered in \cite{horowitz}), ${\tilde {\mu }}=0$
and only the $F^{-}_{\omega lm }$ mode vanishes
at infinity.

\section{Boundary conditions and basis modes}
\label{sec:boundary}

In this section we shall construct sets of basis modes
for our classical scalar field, subject to appropriate
boundary conditions at the event horizon and infinity.
By ``basis modes'', we mean that the general classical
solution of the field equation (\ref{eq:KGone}) 
at each point in the geometry can be
written as a linear combination of the sets of the modes
we shall give below. 
Given that anti-de Sitter space is non-globally hyperbolic,
we do not address the more difficult question of
whether providing initial conditions for the field
on some surface will uniquely control the evolution of the
field through the rest of the space-time.
We shall return to this question in a subsequent article
when we consider quantum states on KN-AdS.

The analysis of the field modes near the event horizon 
is straightforward.  
All modes have the form
\begin{equation}
  f_{\omega lm }(R) = A_{\omega lm} e^{i\Omega R}
    +B_{\omega lm } e^{-i\Omega R} ,
\label{eq:hor}
\end{equation}
with the component proportional to $A_{\omega lm }$ 
corresponding to flux 
coming up out of the past event horizon
${\cal {H}}^{-}$ and the component proportional to 
$B_{\omega lm }$ corresponding to flux 
going down the future event horizon ${\cal {H}}^{+}$.

The key factor in any study of fields on anti-de Sitter space
is the choice of boundary conditions at infinity. 
To understand the options, we first write the general solution
to the radial equation (\ref{eq:Rradial}) 
as $R\rightarrow \infty $ as follows:
\begin{list}{}{}
\item {\em {Case 1}}: ${\tilde {\mu }}^{2} < -9/4l^{2}$
  \begin{equation}
    f_{\omega lm }(R) = C_{\omega lm } e^{iqR} +
      D_{\omega lm } e^{-iqR} ,
      \label{eq:inf1}
  \end{equation}
\item {\em {Case 2}}: ${\tilde {\mu }}^{2} > -9/4l^{2}$
  \begin{equation}
    f_{\omega lm }(R) = 
      \frac {1}{{\sqrt {2}}} C_{\omega lm } \left(
      e^{qR}-ie^{-qR} \right) + 
      \frac {1}{{\sqrt {2}}} D_{\omega lm }
      \left( e^{qR}+ie^{-qR} \right) ,
      \label{eq:inf2}
  \end{equation}
\end{list}
where, in both cases, $q$ is real and positive, and
$C_{\omega lm }$ and $D_{\omega lm }$ are complex constants.

In order to interpret these modes, we calculate the
radial flux at infinity in each case.
Since we are considering behaviour at infinity, the 
appropriate co-ordinates are $(T,y,\Theta ,\Phi )$ and
the radial flux is given by
\begin{displaymath}
  j^{y}  = - i g^{yy} \left( {\bar {\Psi }} \partial _{y} \Psi
    -\Psi \partial _{y} {\bar {\Psi }} \right) , 
\end{displaymath}
where we use a bar to denote complex conjugate.
For our separable wave modes (\ref{eq:modes}),
using the chain rule, the reality of the angular functions
$G_{\omega lm }(\theta )$ and the form of the radial
function $F_{\omega lm }(r)$ (\ref{eq:fdef}), 
the radial flux is now
\begin{displaymath}
  j^{y} = -
  \frac {ir(r^{2}+a^{2})}{\rho ^{2}y} 
    \left( 1+\frac {y^{2}}{l^{2}} \right) 
    G_{\omega lm }^{2}(\theta )
    \Delta _{r}^{-1} \left( {\bar {f}}(R) f'(R) -
      f(R) {\bar {f}}'(R) \right) . 
\end{displaymath}
Therefore, we need to calculate 
${\bar {f}}(R) f'(R) - f(R) {\bar {f}}'(R)$ 
for the two cases given above.
The result is identical for the two cases:
\begin{displaymath}
{\bar {f}}(R) f'(R) - f(R) {\bar {f}}'(R) =
2iq \left( |C_{\omega lm }|^{2} - |D_{\omega lm }|^{2} \right) .
\end{displaymath}
An outgoing flux of particles will correspond to a positive
$j^{y}$, so, in each case, the modes with coefficients $C_{\omega lm}$
represent outgoing flux, while those with coefficients $D_{\omega lm}$
represent incoming flux.  
This is in accord with our expectations for those modes
which have a wave-like form $e^{\pm iqR}$ at infinity.
However, we have shown that linear combinations of
those modes with behaviour $e^{\pm qR}$ at infinity also
have an incoming/outgoing wave interpretation.
This is confirmed by calculating the flux of energy and
angular momentum for the modes at infinity.
In each case, the $C_{\omega lm }$ parts give an outgoing
flux of both energy and angular momentum, while the $D_{\omega lm }$
parts represent incoming energy and angular momentum.

The crucial first step in studying quantum field theory 
in curved space is the definition of appropriate 
basis field modes.
Our goal in the present article is the construction of 
such modes.
This depends critically on the boundary conditions we
impose on the field at infinity.
On anti-de Sitter space, a consistent
quantum field theory can be formulated
for two different boundary conditions on the modes
at infinity \cite{avis}, namely
``transparent'' and ``reflective''.
Ref. \cite{avis} considers only a conformally coupled
scalar field, but the concepts apply equally well to the
general coupling we are considering here.

``Reflective'' boundary conditions correspond to an
absence of flux across infinity.
In our situation, this means that we must have
$|C_{\omega lm }|=|D_{\omega lm }|$ in either
(\ref{eq:inf1}) or (\ref{eq:inf2}), as applicable.
Linearly independent sets of basis modes may 
be constructed by taking, firstly $C_{\omega lm }=D_{\omega lm}$
and secondly $C_{\omega lm }=-D_{\omega lm }$.
Thus there are two sets of modes, corresponding to this choice
of sign for $C_{\omega lm }$.
This is because there are two boundary conditions which
lead to zero flux across infinity: 
Dirichlet (in which the field itself vanishes at the boundary)
and Neumann (where the derivative of the field vanishes).

The natural basis modes in this case are therefore
defined by, in case 1,
\begin{eqnarray*}
  f_{\omega lm }^{1}(R) & \sim & \left\{
      \begin{array}{ll}
         e^{i\Omega R}
        +B_{\omega lm }^{1} e^{-i\Omega R}
        & {\mbox{as $R\rightarrow -\infty $,}} \\
        {\tilde {C}}_{\omega lm } \cos (qR)
        & {\mbox {as $R \rightarrow \infty $,}} 
      \end{array}
      \right.
\\
  f_{\omega lm}^{2}(R) & \sim & \left\{
      \begin{array}{ll}
         e^{i\Omega R}
        +B_{\omega lm }^{2} e^{-i\Omega R}
        & {\mbox{as $R\rightarrow -\infty $,}} \\
         -{\tilde {D}}_{\omega lm} \sin (qR)
        & {\mbox {as $R \rightarrow \infty $,}}
      \end{array}
      \right.
\end{eqnarray*}
and, in case 2,
\begin{eqnarray*}
  f_{\omega lm }^{1}(R) & \sim & \left\{
      \begin{array}{ll}
         e^{i\Omega R}
        +B_{\omega lm }^{1} e^{-i\Omega R}
        & {\mbox{as $R\rightarrow -\infty $,}} \\
        \frac {1}{{\sqrt {2}}}{\tilde {C}}_{\omega lm } e^{qR}
        & {\mbox {as $R \rightarrow \infty $,}} 
      \end{array}
      \right.
\\
  f_{\omega lm}^{2}(R) & \sim & \left\{
      \begin{array}{ll}
         e^{i\Omega R}
        +B_{\omega lm }^{2} e^{-i\Omega R}
        & {\mbox{as $R\rightarrow -\infty $,}} \\
        \frac {1}{{\sqrt {2}}}{\tilde {D}}_{\omega lm} e^{-qR}
        & {\mbox {as $R \rightarrow \infty $.}}
      \end{array}
      \right.
\end{eqnarray*}
In both cases 1 and 2, the modes $f_{\omega lm }^{1}$ arise
from the choice $C_{\omega lm }=D_{\omega lm}$
(with ${\tilde {C}}_{\omega lm }=2C_{\omega lm}$) and
the modes $f_{\omega lm }^{2}$ from $C_{\omega lm}=-D_{\omega lm }$
(with ${\tilde {D}}_{\omega lm}=-2iC_{\omega lm}$).

In each case, both sets of modes $f_{\omega lm }^{1}$ and 
$f_{\omega lm }^{2}$ represent unit flux coming up out of the 
past event horizon ${\cal {H}}^{-}$, being completely reflected
at infinity ${\cal {I}}$ and going back down the future
event horizon ${\cal {H}}^{+}$.
Note that we do not rule out a change of phase on reflection,
so $B_{\omega lm}^{i}$ does not necessarily equal 
$1$ for $i=1,2$.
These modes are sketched on the Penrose diagram of KN-AdS in
figure 2.
\begin{figure}
\begin{center}
\includegraphics[height=3in]{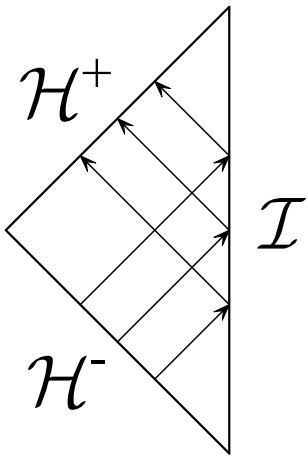}
\caption{Basis modes on KN-AdS defined by ``reflective'' boundary
conditions at infinity. Flux comes out of the past horizon
${\cal {H}}^{-}$, is entirely reflected at ${\cal {I}}$ and
returns down the future horizon ${\cal {H}}^{+}$.}
\label{figure2}
\end{center}
\end{figure}

``Transparent'' boundary conditions allow flux
to escape through time-like infinity.
These boundary conditions allow
the flow of information through time-like infinity,
which needs to be carefully considered if one
is interested in quantum states on this background.
However, our purposes here are purely classical,
and, furthermore, we do not address the 
difficult issue of how the field evolves in time,
and whether such evolution can be fixed by some
initial data (given that anti-de Sitter space
does not possess a global Cauchy surface).
At the same time, these 
``transparent'' boundary conditions are the analogue for
asymptotically anti-de Sitter black holes of the usual 
behaviour of basis field modes on asymptotically
flat space-times (where, in general, particles 
escape to null infinity),
and so may be most pertinent if when we subsequently
come to consider the construction of appropriate
quantum states on KN-AdS.
We therefore follow the practice in Kerr space-time
\cite{ottewill} and define ``in'' and ``up'' modes
as follows: in case 1,
\begin{eqnarray*}
  f_{\omega lm}^{in} & \sim & \left\{
    \begin{array}{ll}
      B^{in}_{\omega lm} e^{-i\Omega R} &
      {\mbox {as $R\rightarrow -\infty $,}} \\
      e^{-iqR}+C_{\omega lm}^{in} e^{iqR} &
      {\mbox {as $R\rightarrow \infty $,}}
    \end{array} \right.
\\
  f_{\omega lm }^{up} & \sim & \left\{
    \begin{array}{ll}
      e^{i\Omega R} + B^{up}_{\omega lm} e^{-i\Omega R} &
      {\mbox {as $R\rightarrow -\infty $,}} \\
      C_{\omega lm }^{up} e^{iqR} &
      {\mbox {as $R\rightarrow \infty $,}}
    \end{array} \right.
\end{eqnarray*}
and in case 2,
\begin{eqnarray*}
  f_{\omega lm}^{in} & \sim & \left\{ 
    \begin{array}{ll}
      B^{in}_{\omega lm} e^{-i\Omega R} &
      {\mbox {as $R\rightarrow -\infty $,}} \\
      \frac {1}{{\sqrt {2}}}(e^{qR}+ie^{-qR})
      +\frac {1}{{\sqrt {2}}} C_{\omega lm}^{in} 
      (e^{qR}-ie^{-qR}) &
      {\mbox {as $R\rightarrow \infty $,}}
    \end{array} \right.
\\
  f_{\omega lm }^{up} & \sim & \left\{
    \begin{array}{ll}
      e^{i\Omega R} + B^{up}_{\omega lm} e^{-i\Omega R} &
      {\mbox {as $R\rightarrow -\infty $,}} \\
      \frac {1}{{\sqrt {2}}} C_{\omega lm }^{up} 
      (e^{qR} -ie^{-qR}) &
      {\mbox {as $R\rightarrow \infty $,}}
    \end{array} \right.
\end{eqnarray*}
In both cases, the $f_{\omega lm}^{in}$ modes represent
unit incoming flux from infinity, which is partly reflected back
to infinity and partly goes down the future event horizon 
${\cal {H}}^{+}$.
The $f_{\omega lm}^{up}$ modes correspond to unit flux
coming up from the past event horizon ${\cal {H}}^{-}$,
which is partially reflected back down the future event horizon
${\cal {H}}^{+}$ and partially transmitted to infinity.
These modes are sketched in figure 3 (compare \cite{ottewill,frolov}
for the corresponding modes in Kerr-Newman space-time).
\begin{figure}
\begin{center}
\includegraphics[height=3in]{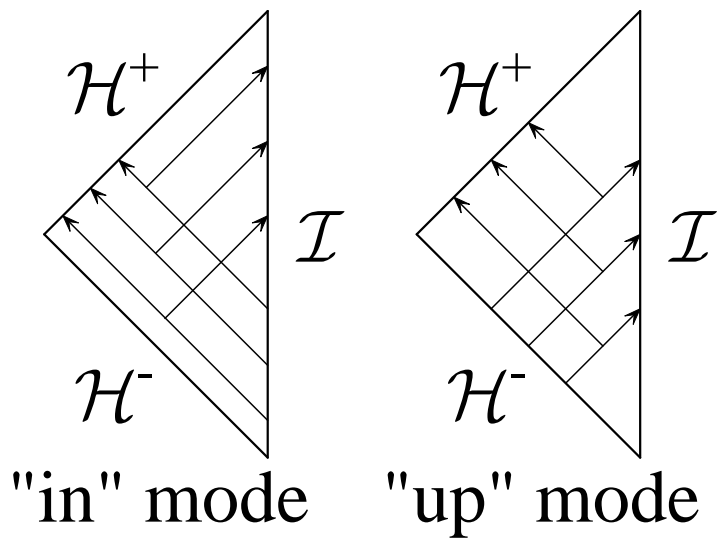}
\caption{Basis modes on KN-AdS defined by ``transparent'' boundary
conditions at infinity. For the ``in'' modes, 
flux comes in from infinity, is partly reflected back to 
infinity, and partly goes down the future horizon ${\cal {H}}^{+}$.
For the ``up'' modes, flux comes up from the past horizon
${\cal {H}}^{-}$, is partly reflected down 
the future horizon ${\cal {H}}^{+}$ and partly escapes to infinity.}
\label{figure3}
\end{center}
\end{figure}

\section{Definition of positive frequency and classical super-radiance}
\label{sec:super}

We are now in a position to discuss whether classical
scalar field modes in KN-AdS exhibit super-radiance 
phenomena.
For the moment we consider a very general field mode, having 
behaviour at the event horizon given by (\ref{eq:hor}),
and whose form at infinity is given by either (\ref{eq:inf1}) or
(\ref{eq:inf2}), as applicable.
If $f_{1}$ and $f_{2}$ are solutions of the radial equation
(\ref{eq:Rradial}), then the Wronskian
\begin{displaymath}
f_{1}'(R) {\bar {f}}_{2} (R) - {\bar {f}}_{2}'(R) f_{1}(R)
\end{displaymath}
will be constant.
Setting $f_{2}=f_{1}=f_{\omega lm}$, we arrive at the
relation:
\begin{equation}
  \Omega \left( \left| A_{\omega lm } \right| ^{2} 
  -\left| B_{\omega lm } \right| ^{2} \right)
  = q \left( \left| C_{\omega lm} \right| ^{2} 
  -\left| D_{\omega lm} \right| ^{2} \right) .
\label{eq:wronskian}
\end{equation}
We should stress at this stage, that, in contrast to the
corresponding equation for scalar field modes in Kerr-Newman
space-time \cite{dewitt}, here $q$ is by definition always
positive and, furthermore, it is a constant for all
field modes, depending only on the coupling constants
of our original model (see section \ref{sec:modes}).
We now apply equation (\ref{eq:wronskian}) 
to each of our sets of basis modes.

Firstly, for modes defined by ``reflective'' boundary
conditions at infinity, $|C_{\omega lm}|=|D_{\omega lm }|$,
so the right hand side of (\ref{eq:wronskian}) vanishes.
Thus, for these modes (which were defined with $A_{\omega lm }=1$),
it must be the case that $|B_{\omega lm }|=1$.
This is in agreement with conservation of flux, as 
there is no flux of particles across ${\cal {I}}$,
so the flux emitted from ${\cal {H}}^{-}$ is 
entirely reflected back down ${\cal {H}}^{+}$.
Note that it is not necessarily the case that $B_{\omega lm }=1$
as there may be a phase shift on reflection at infinity.
We conclude that, for a field subject to ``reflective''
boundary conditions at infinity, there is no classical
super-radiance.
This is in complete accord with the energy argument of
Ref. \cite{hr}, who showed that for matter satisfying the
dominant energy condition, and ``reflective'' boundary conditions,
there has to be a net flux of energy down the event horizon.

Secondly, we consider modes defined by ``transparent''
boundary conditions.
For ``in'' modes, $A_{\omega lm}^{in}=0$,
$D_{\omega lm }^{in}=1$ and the 
relation (\ref{eq:wronskian}) takes the form:
\begin{displaymath}
  -\Omega \left| B_{\omega lm }^{in} \right| ^{2} 
  = q \left( \left| C_{\omega lm}^{in} \right| ^{2} 
  -1 \right) .
\end{displaymath}
Therefore, for modes with $\Omega <0$, we have
$|C_{\omega lm}^{in}|>1$, in other words,
the flux transmitted out to infinity is greater than
the flux incoming from infinity.
Similarly, for ``up'' modes, $D_{\omega lm}^{up}=0$
and $A_{\omega lm }^{up}=1$,
giving
\begin{displaymath}
  \Omega \left( 1 
  -\left| B_{\omega lm }^{up} \right| ^{2} \right)
  = q \left| C_{\omega lm}^{up} \right| ^{2} .
\end{displaymath}
In this case we have $|B_{\omega lm }^{up}|>1$
for modes with $\Omega <0$, that is, the flux 
reflected back down ${\cal {H}}^{+}$ is greater than
that coming up out of ${\cal {H}}^{-}$.
In conclusion, for modes defined with ``transparent''
boundary conditions at infinity, we have classical
super-radiance if $\Omega <0$.

However, we need at this stage to address the 
question of whether $\Omega $ can be negative
in practice.
This is closely tied to the definition of
positive frequency (which we have side-stepped
until now).  
For rotating black hole geometries, the
question of positive frequency is a rather thorny
one, and its resolution is one of the 
reasons that quantum field theory in Kerr-Newman
black holes is so much more complex than
for Schwarzschild (or Reissner-Nordstr\"om) black holes
(see, for example, \cite{ottewill} and references therein
for some discussion of this problem).
The crux of the problem is the lack of a globally
time-like Killing vector, which means that 
a decision has to be made on the location
in the geometry of a locally stationary observer
who we wish to see only positive frequency modes.
In Kerr-Newman, the convention is to take ``in'' modes
to have positive frequency with respect to a locally
non-rotating observer (LNRO, also referred to as a 
zero angular momentum observer (ZAMO) by some authors,
for example, \cite{bardeen} or a fiducial observer
(FIDO) in other texts, such as \cite{membrane}) 
at infinity (the ``distant observer'' viewpoint \cite{frolov}),
and ``up'' modes to 
have positive frequency as measured by an LNRO close
to the event horizon (the ``near-horizon observer'' viewpoint
\cite{frolov}).
If we apply this definition of positive frequency in our
case, then the time-like Killing vector for an LNRO at 
infinity is $\partial /\partial T$ rather than
$\partial /\partial t$.
Therefore we require our ``in'' modes 
$\Psi _{\omega lm }^{in}$ to be such that
\begin{displaymath}
  \frac {\partial }{\partial T} 
  \Psi _{\omega lm }^{in} = W \Psi _{\omega lm }^{in}
\end{displaymath}
where $W$ should be positive.
Using the separation of the field modes (\ref{eq:modes})
we find that
\begin{displaymath}
   W = \omega + \frac {am}{l^{2}} >0.
\end{displaymath}
Therefore it is possible, with this scheme, to have
``in'' modes for which $W>0$ but $\Omega <0$, and
therefore these modes will exhibit super-radiance.
For the ``up'' modes, it is conventional to take 
${\tilde {\omega }}$ (see equation (\ref{eq:tildeomega}))
to be positive. 
In Kerr-Newman space-time, there are super-radiant ``up''
modes when $\omega <0$ but ${\tilde {\omega }}>0$.
However, in this case there are no super-radiant ``up'' modes
if we define ${\tilde {\omega }}>0$, as $q$ is always
positive.
This reveals how the presence or absence of super-radiant modes
is dependent in KN-AdS on the definition of positive
frequency, whereas for Kerr-Newman black holes super-radiant
modes are inevitable whatever the choice of positive frequency.
Indeed, in KN-AdS we could reasonably define positive frequency
for ``in'' modes with respect to an observer near the event horizon
rather than an observer at infinity.
In this case there would be no super-radiant ``in'' modes either.
However, for general KN-AdS black holes it is by no means obvious
from the physical point of view that this is a reasonable strategy.

On the other hand, for a certain class of KN-AdS black holes
there is a globally time-like Killing vector.
As outlined in section \ref{sec:geom}, providing the 
angular velocity of the event horizon in $(T,y,\Theta , \Phi )$
co-ordinates does not exceed $l^{-2}$, the Killing vector
$\zeta $ which is null on the event horizon is time-like 
everywhere outside the event horizon.
In this situation the question of how to define positive
frequency has a natural resolution: we require all 
field modes $\Psi _{\omega lm }$ to be such that
$\zeta \Psi _{\omega lm }={\tilde {W}} \Psi _{\omega lm }$ where
${\tilde {W}}>0$.
Direct calculation reveals that ${\tilde {W}}$ is in fact
${\tilde {\omega }}$, so we require that ${\tilde {\omega }}$
and hence $\Omega $ (see equation (\ref{eq:Omegadef}))
be positive for all field modes.
Therefore there are no super-radiant modes for these KN-AdS
black holes.

We may summarize the results of this section as follows.
The presence or absence of super-radiant modes in KN-AdS
depends on two factors: the boundary conditions at infinity
and our definition of positive frequency modes.
There are two possible boundary conditions at infinity.
Firstly, ``reflective'' boundary conditions lead to 
modes which exhibit no super-radiance, regardless of
our definition of positive frequency or the rotation 
of the black hole.
For ``transparent'' boundary conditions, there is a
choice of positive frequency which rules out
super-radiant modes.
This choice is the natural one for KN-AdS geometries
having a globally time-like Killing vector.
However, if there is no globally time-like Killing vector,
then the choice of positive frequency is ambiguous 
and super-radiance is possible.

\section{Conclusions}
\label{sec:conc}

The main result of this article is that classical
super-radiance is not inevitable for KN-AdS black holes,
unlike the situation for rotating black holes in asymptotically
flat space.
This result is a little surprising given that all KN-AdS black 
holes have an ergosphere, 
although energy considerations \cite{hr} have suggested
that those KN-AdS black holes
possessing a globally time-like Killing vector are classically stable.

There are two key properties of classical field modes in KN-AdS
as far as super-radiance is concerned: 
the boundary conditions at infinity and the definition 
of positive frequency.
Both these factors are anticipated to be important in quantum
field theory studies.
We have shown that using ``reflective'' boundary conditions
(that is, Dirichlet or Neumann boundary conditions) 
at infinity, for all KN-AdS black hole super-radiance is
absent.  
Using ``transparent'' boundary conditions, however, 
the situation depends on the choice of positive frequency.
For those KN-AdS black holes having a globally time-like
Killing vector, the natural definition of positive 
frequency implies that there are no super-radiant modes.
For other KN-AdS black holes, there is a choice of
positive frequency for which super-radiance is absent.
This is to be contrasted with the situation for
Kerr-Newman black holes in asymptotically flat space,
where super-radiance occurs regardless of the definition
of positive frequency.

Along the way, we have constructed sets of basis field modes
(for both boundary conditions at infinity) which will be
useful for studying quantum field theory on KN-AdS.
This will be the subject of further work.
However, we are encouraged by the results of this paper,
since the absence of super-radiance for the cases outlined
above removes one of the difficulties of studying quantum 
field theory on rotating black hole geometries.
The next step will be to investigate whether the quantum 
analogue of super-radiance \cite{unruh} is present for fields
on KN-AdS, given that, for asymptotically flat Kerr-Newman black
holes,
fermion fields do not classically super-radiate but the
quantum spontaneous emission process does occur.  

One might hope that quantum field theory will be more 
straightforward on those
KN-AdS black holes having a globally time-like
Killing vector, since there is a natural 
definition of positive frequency.
Since these black holes do not have a velocity of light surface,
it might even be possible to define a regular Hartle-Hawking state.
It is well known that ordinary Kerr-Newman black holes
do not possess a Hartle-Hawking state \cite{kay},
and it is an elementary consequence of the conservation equations
that any thermal state which rotates rigidly with the angular
velocity of the event horizon must be divergent at the
velocity of light surface \cite{ottewill1}.
This proof breaks down in the absence of a velocity of light surface,
although showing that a Hartle-Hawking state can be defined
will not be a straightforward problem.
In the current paper, we have not addressed the difficult
question, fundamental to the construction of quantum states,
of how to define appropriate initial data for the field, given
that anti-de Sitter space is not globally-hyperbolic.
Studies on Kerr black holes \cite{ottewill} have revealed
some of the difficulties of trying to define quantum states
by the behaviour of the field on a surface which is not a Cauchy 
surface for the geometry (as happens, for example, when attempting
to define the Hartle-Hawking state).  
These subtleties are anticipated to be more complex when dealing
with asymptotically anti-de Sitter black holes, and will be 
studied in detail in the future.

\section*{Acknowledgements}
We would like to thank Robert Mann, Vitor Cardoso, Jos\'e Lemos and
Bin Wang for helpful comments and bringing a number of references
to our attention.

\end{document}